\documentclass[a4paper,11pt]{article}

\usepackage{bm}
\usepackage{latexsym}
\usepackage{amssymb}
\usepackage{amsmath}
\usepackage{mathtools}
\usepackage{hyperref}
\hypersetup{colorlinks=true, allcolors=blue}
\usepackage{graphicx}

\usepackage{mathbbol}
\usepackage{braket}
\usepackage{empheq}
\usepackage{cite}
\usepackage[title]{appendix}
\usepackage{here}
\usepackage{pifont}
\usepackage{comment}

\makeatletter
\renewcommand\subparagraph{
    \@startsection {subparagraph}{5}{\z@ }{3.25ex \@plus 1ex
    \@minus .2ex}{-1em}{\normalfont \normalsize \bfseries }}
\makeatother

\numberwithin{equation}{section}

\begin{document}
\pagestyle{empty}

\vspace{-4cm}
\begin{center}
    \hfill KEK-TH-2827, YITP-26-44 \\
\end{center}

\vspace{2cm}

\begin{center}

{\bf\LARGE
Super-Heisenberg protocol for dark matter and high-frequency gravitational wave search \\
}

\vspace*{1.5cm}
{\large 
Wakutaka Nakano$^1$ and Ryoto Takai$^{2,3,4}$
} \\
\vspace*{0.5cm}

{\it 
$^1$Department of Physics, Tokyo Metropolitan University, Minami-Osawa, Hachioji-shi, Tokyo 192-0397, Japan \\
$^2$KEK Theory Center, Tsukuba 305-0801, Japan \\
$^3$The Graduate University for Advanced Studies (SOKENDAI), Tsukuba 305-0801, Japan\\
$^4$Yukawa Institute for Theoretical Physics, Kyoto University, Kyoto 606-8502, Japan \\
}

\end{center}

\vspace*{1.0cm}

\begin{abstract}
{\normalsize \noindent
We propose a quantum-enhanced sensing scheme for the detection of
wave-like dark matter and high-frequency gravitational waves using
two-dimensional ion crystals in a Penning trap.
The protocol employs spin-motion squeezed states to improve the
signal-to-noise ratio and enable a super-Heisenberg scaling with
respect to the number of ions over a broad parameter range.
We analyze the sensitivity of the protocol to representative wave-like
dark matter candidates, including the axion-like particle and the dark
photon, as well as to high-frequency gravitational waves, taking into
account the decoherence and dephasing of the ion spins. Our results
indicate that two-dimensional ion crystals and this new protocol
provide a promising platform for probing previously unexplored
parameter space in searches for light dark matter and high-frequency
gravitational waves.
}
\end{abstract} 


\newpage
\baselineskip=18pt
\setcounter{page}{2}
\pagestyle{plain}

\setcounter{footnote}{0}

\tableofcontents
\noindent\hrulefill


\section{Introduction}

Quantum metrology provides a powerful framework for detecting extremely
weak signals by exploiting nonclassical resources such as squeezing and
entanglement.
A variety of platforms have been explored for quantum
sensing~\cite{Degen:2016pxo}.
Among them, trapped-ion systems offer a particularly well-controlled
setting, combining long coherence times with high-fidelity control of
internal states and precise manipulation of collective motional
modes~\cite{Leibfried:2003zz,Blatt:2008jzf}.
In large ion crystals, for instance, spin-dependent optical dipole forces
couple the collective spin with the motion~\cite{Britton:2012krt}, which
has been exploited in recent experiments to demonstrate quantum-enhanced
sensing of weak displacements and electric fields~\cite{Gilmore:2021qqo}.

In conventional protocols with $N$ qubits, the achievable precision is
bounded by the standard quantum limit $N^{-1/2}$ for uncorrelated
probes and can approach the Heisenberg limit $N^{-1}$ when entangled
states are employed~\cite{Braunstein:1994zz}.
While it has been recognized that nonlinear parameter encoding and
many-body interactions can yield sensitivities that scale more favorably
than the Heisenberg scaling, referred to as super-Heisenberg
scaling~\cite{Boixo:2006zvy,Boixo:2008gxa,Napolitano:2010hal}, the
interpretation of such scaling depends sensitively on the definition of
the resources being counted, requiring care in comparing different
approaches~\cite{Rams:2018jor}.
Recently, it has been shown that suitably engineered spin-boson dynamics
can give rise to a super-Heisenberg scaling of the quantum Fisher
information~\cite{Pavlov:2024uxs,Pavlov:2026zcj}.

Understanding the nature of dark matter is one of the most important open
problems in modern physics.
If dark matter is sufficiently light, it can behave as a coherently oscillating classical field in the Galactic
halo~\cite{Arias:2012az,Graham:2013gfa}.
Representative candidates for such wave-like dark matter include axion-like
particles and dark photons.
A variety of quantum sensing approaches have been developed to search for
light dark matter, offering complementary strategies across different
experimental platforms~\cite{Dixit:2020ymh,Chigusa:2024psk,Chen:2023swh,
Filzinger:2023qqh,Jiang:2023jhl,AION:2025nio,
Chigusa:2025rqs,Kang:2025kaf}, including trapped-ion
systems~\cite{Gilmore:2021qqo,Ito:2023zhp,Badurina:2025idj}.
These oscillatory backgrounds can induce weak effective electric fields
that drive collective motional modes of ions.
Under resonant driving conditions, the resulting motional displacement can
be mapped onto a measurable spin signals.

A similar detection principle also applies to high-frequency gravitational
waves.
While most searches have traditionally focused on the Hz--kHz frequency
band~\cite{KAGRA:2013rdx,LIGOScientific:2016aoc}, various scenarios predict
signals at much higher frequencies~\cite{Aggarwal:2025noe}.
This has stimulated the development of detector concepts that go beyond
conventional interferometric approaches~\cite{Goryachev:2014yra,
chou2017mhz,Carney:2024zzk,AION:2025igp}.
In ion traps, such high-frequency gravitational waves can couple to the
ions and resonantly excite its oscillation
modes~\cite{Takai:2025cyy,Ito:2025mgm}.

In this work, we study the application of a super-Heisenberg protocol to
wave-like dark matter and high-frequency gravitational wave search by
improving the protocol with two-dimensional ion crystals in
Refs.~\cite{Gilmore:2021qqo,Ito:2025mgm}.
Ions confined in a Penning trap self-organize into a planar crystal due
to the balance between the trap potential and Coulomb repulsion,
supporting collective vibrational modes.
These modes can be resonantly excited by dark matter or gravitational waves.
The resulting axial motion is transduced into a measurable collective-spin signal via optical dipole forces.
We show that a protocol based on spin-motion squeezed states enables
sensitivities that surpass the Heisenberg scaling, thereby enabling
exploration of previously inaccessible regions of parameter space.


\section{Experimental setup}

In this section, we introduce the physical setup and outline the sensing
protocol employed in this work.
We begin in Sec.~\ref{sec:ioncrystal} with a brief review of
two-dimensional ion crystals in a Penning trap and the optical dipole
forces which connect the spins and the motional degrees of freedom.
In Sec.~\ref{sec:state}, we introduce the spin-motion squeezed state that
serve as a key resource for quantum-enhanced sensing.
Building on these elements, Sec.~\ref{sec:protocol} presents our protocol,
which combines the spin-motion squeezed state with the idea of
Refs.~\cite{Gilmore:2021qqo,Ito:2025mgm} to achieve enhanced sensitivity
beyond the Heisenberg scaling.


\subsection{Two-dimensional ion crystals}
\label{sec:ioncrystal}

We consider a two-dimensional ion crystal confined in a Penning trap,
where $N$ ions are subject to a combination of a quadrupole electric field
and a strong magnetic field~\cite{Sawyer:2012rjk,Britton:2012krt}.
In the rotating frame co-moving with the ions, the effective trap potential
is harmonic in both axial and radial directions, ensuring stable
confinement under appropriate conditions on the rotation frequency.
Multiple ions self-organize into a triangular Wigner crystal in the plane
perpendicular to the magnetic field, with equilibrium positions determined
by the balance between the trapping potential and Coulomb repulsion.

The axial motion of the ions is well described within the harmonic
approximation~\cite{huang1998precise,porras2006quantum}, where the
potential with respect to the axial motion
\begin{align}
    V_z &= \frac{1}{2} m_{\rm ion} \omega_z^2 \sum_{i=1}^N z_i^2 +
    \frac{1}{2} \sum_{i \neq j} \frac{\alpha_{\rm EM}} {\sqrt{\lvert
    {\bm \rho}_i - {\bm \rho}_j \rvert^2 + (z_i - z_j)^2}}  \\
    &= \frac{1}{2} m_{\rm ion} \omega_z^2 \sum_i z_i^2 -
    \frac{\alpha_{\rm EM}}{4} \sum_{i \neq j} \frac{(z_i - z_j)^2}{\lvert
    {\bm \rho}_i - {\bm \rho}_j \rvert^3} + \cdots ,
\end{align}
gives rise to a set of collective phonon modes, with $m_{\rm ion}$,
$\omega_z$, $\alpha_{\rm EM}$, and $({\bm \rho}_i, z_i)$ denoting the ion
mass, the angular frequency of the single-ion axial mode, the
fine-structure constant, and the position of the $i$th ion, respectively.
The corresponding Hamiltonian can be diagonalized as $H_0 = \sum_k \omega_k
\hat{a}^\dagger_k \hat{a}_k$, leading to a spectrum of axial vibrations
often referred to as drumhead modes in the large-system limit,
\begin{equation}
    \hat{z}_i = \sum_k \frac{b_{ik}}{\sqrt{2 m_{\rm ion} \omega_k}}
    \left( \hat{a}_k e^{-i \omega_k t} + \hat{a}^\dagger_k
    e^{i \omega_k t} \right),
\end{equation}
where $b_{ik}$ is the orthonormal matrix diagonalizing $V_z$, $\omega_k$
is the frequency for the $k$th mode, and $\hat{a}_k^{(\dagger)}$ is the
ladder operator.
One can find the explicit formulae of $b_{ik}$ in the large-system limit
in Ref.~\cite{Ito:2025mgm}.
The center-of-mass mode, in which all ions oscillate in phase, has the
highest frequency, while lower-frequency modes exhibit nontrivial spatial
structures and may be nearly degenerate depending on the crystal geometry.
In addition to motional degrees of freedom, each ion hosts a qubit
system encoded in internal states, enabling coherent spin manipulation
and state readout.
In this work, we focus on the use of $^9$Be$^+$ ions, but other species
are available with the same protocol.

Spin-motion coupling is realized by optical dipole forces, generated by
the interference of two laser beams applied to the ion
crystal~\cite{Britton:2012krt,Gilmore:2021qqo}.
By appropriately choosing the laser polarizations, the differential AC
Stark shift between the spin states can be canceled, leaving a spatially
modulated, state-dependent force.
In the Lamb--Dicke regime, this results in a force proportional to the
axial displacement and oscillating at the beat frequency of the lasers.
The resulting interaction couples the collective spin to the phonon modes
and enables the transduction of motional excitations into measurable spin
dynamics.

The spatially homogeneous optical dipole force predominantly couples to
the center-of-mass mode, and one can engineer spatially inhomogeneous
optical dipole force profiles by modulating the phase of the laser fields
across the crystal~\cite{Polloreno:2022nxl}.
By tuning the drive frequency to a target phonon mode and shaping the
spatial dependence appropriately, the optical dipole force can selectively
couple to specific collective modes, including nearly degenerate
excitations.
This tunable spin-phonon interaction
\begin{equation}
    \hat{H}_{\rm ODF} = \frac{g}{\sqrt{N}} \left( \hat{a} + \hat{a}^\dagger
    \right) \hat{J}_z ,
    \label{eq:odf}
\end{equation}
where $\hat{a}^{(\dagger)}$ and $\hat{J}_z$ denote the ladder operator of
the mode in focus and the $z$ component of the total spin, respectively,
provides a key resource for mode-selective sensing and for generating
nonclassical states relevant to quantum-enhanced metrology.


\subsection{Spin-motion squeezed state}
\label{sec:state}

The Heisenberg limit sets the ultimate bound on the precision of parameter
estimation when the parameter is encoded via a collective unitary
evolution acting on $N$ quantum resources, as opposed to independent
or separable operations.
In the protocol introduced below, both the $N$ spins and the motional
degrees of freedom are utilized, thereby circumventing the
assumptions underlying the Heisenberg scaling, where the
spin-motion squeezed state
\begin{equation}
    \Ket{{\rm init}} = \sum_{n=0}^J c_{2n} \Ket{-J + 2n, 0} +
    \sum_{n=1}^\infty c_{2n+2J} \Ket{J, 2n}
    \label{eq:init}
\end{equation}
plays a central role.
Here, the state $\Ket{m, n}$ denotes the simultaneous eigenstate of the
operators $\hat{J}_z$ and $\hat{a}^\dagger \hat{a}$ with eigenvalues $m$
and $n$, respectively, and $J = N / 2$ is the maximum of $m$.

We prepare this state by following Ref.~\cite{Pavlov:2024uxs}.
We start from the state $\Ket{-J, 0}$, and the motional squeezed state
\begin{equation}
    \Ket{\psi} = \sum_{n=0}^\infty a_{2n} \Ket{-J, 2n},
    \quad a_{2n} = \frac{\sqrt{(2n)!}}{2^n \, n!} z^{n/2} (1-z)^{1/4}
    e^{-i n \phi}
\end{equation}
with $z = \tanh^2 r$ is generated using two sideband transitions and
laser cooling~\cite{Kienzler:2015ojd}.
Next, the Tavis--Cummings Hamiltonian
\begin{equation}
    \hat{H}_{\rm TC} (t) = \zeta (t) \, \hat{J}_z +
    \frac{\lambda (t)}{\sqrt{N}} \left( \hat{J}_+ \hat{a}
    + \hat{J}_- \hat{a}^\dagger \right)
\end{equation}
is performed to $\Ket{\psi}$ adiabatically.
Here, $\hat{J}_\pm = \sum_i \hat{\sigma}^\pm_i = \sum_i (\hat{\sigma}^x_i
\pm i\hat{\sigma}^y_i) / 2$.
At the initial time $t_i$, the detuning $\zeta (t)$ is set to satisfy
$\vert \zeta (t_i) \vert \gg \lambda (t_i)$ and $\zeta (t_i) < 0$, such
that the state $\Ket{-J, 2n}$ are eigenstates of the Tavis--Cummings
Hamiltonian, $\hat{H}_{\rm TC} (t_i) \Ket{-J, 2n} \simeq -J \zeta (t_i)
\Ket{-J, 2n}$.
The detuning function is then adiabatically varied to a final value
satisfying $\zeta (t_f) \gg \lambda (t_f)$ and $\zeta (t_f) > 0$.
Provided that the adiabatic condition holds throughout the evolution, the
system remains in the instantaneous eigenstates of the Hamiltonian and
the conserved operator $\hat{\mathcal{N}} = \hat{a}^\dagger \hat{a} + \hat{J}_z + J$.
Each initial state $\Ket{-J, 2n}$ is adiabatically mapped to the final
state $\Ket{-J + 2n, 0}$ for $n \leq J$ and $\Ket{J, 2n - 2J}$ for
$n > J$.
Consequently, the spin-motion squeezed state~\eqref{eq:init} is obtained
with $c_{2n} = a_{2n} \exp [-i \int^{t_f}_{t_i} {\rm d}t \, E_{2n} (t)]$.


\subsection{Protocol}
\label{sec:protocol}

In this subsection, we demonstrate that phonon-mode excitations can be
detected by following the protocol of Ref.~\cite{Gilmore:2021qqo}, with
the initial state replaced by the spin-motion squeezed state in
Eq.~\eqref{eq:init}, enabling high-sensitivity detection of external
fields such as wave-like dark matter and gravitational waves.
We assume an external wave that resonantly couples to a phonon mode of
the ion crystal, described by the interaction Hamiltonian
\begin{equation}
    \hat{H}_{\rm int} = \alpha \hat{a} + \alpha^* \hat{a}^\dagger ,
    \label{eq:int}
\end{equation}
where $\alpha$ denotes the amplitude of the target field.

First, we prepare the spin-motion squeezed state~\eqref{eq:init}.
After the spin rotation by $\hat{U}_y (\pi / 2) = \exp (-i \frac{\pi}{2}
\hat{J}_y)$, the optical dipole force operation~\eqref{eq:odf} is performed during the
time $\tau$.
We hold the system during the time $T - 2 \tau$, and change the state by
applying the inverse optical dipole force operation.
Then, the spin rotation by $\hat{U}_x (\vartheta) = \exp (-i
\vartheta \hat{J}_x)$ yields the final state
\begin{align}
    \Ket{{\rm fin}} &= \hat{U}_x (\vartheta) e^{-i \tau (\hat{H}_{\rm ODF}
    + \hat{H}_{\rm int})} \hat{U}_x (\pi) e^{-i (T - 2 \tau)
    \hat{H}_{\rm int}} e^{-i \tau (\hat{H}_{\rm ODF} + \hat{H}_{\rm int})}
    \hat{U}_y (\pi / 2) \Ket{{\rm init}} \\
    &= \left[ \hat{U}_x (\vartheta) \hat{U}_z (\varphi) \hat{U}_y
    (\pi / 2) \right] \otimes e^{-i T \hat{H}_{\rm int}} \Ket{{\rm init}}
\end{align}
with
\begin{equation}
    \varphi = \frac{2 g \tau (T - \tau)}{\sqrt{N}} ({\rm Im} \, \alpha) .
    \label{eq:angle} 
\end{equation}
The rotation angle $\vartheta$ is determined to minimize the variance of
$\hat{J}_z$, ${\rm var} (\hat{J}_z) = \langle \hat{J}_z^2 \rangle -
\langle \hat{J}_z \rangle^2$.
Finally, the $z$-component of the total spin $\hat{J}_z$ is measured by
using the spin-dependent global resonance fluorescence
method~\cite{hosten2016quantum,lewis2020protocol}.

\begin{figure}[t!]
    \centering
    \includegraphics[width=0.6\linewidth]{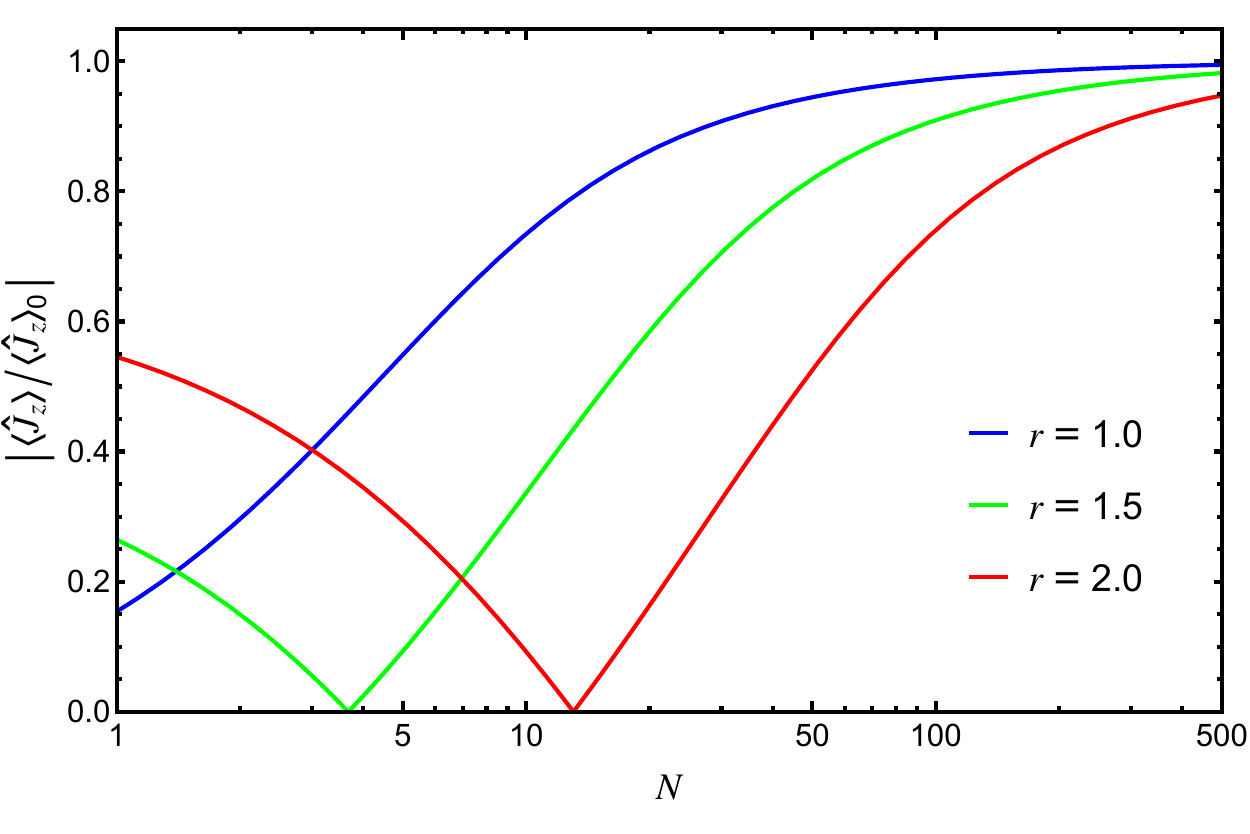}
    \caption{Absolute values of the ratio $\langle \hat{J}_z
    \rangle / \langle \hat{J}_z \rangle_0$ as functions of $N$
    for $r = 1.0$ (blue), $r = 1.5$ (green), and $r = 2.0$ (red), with
    $\vartheta = \pi / 2$.
    Here, $\langle \hat{J}_z \rangle$ and $\langle \hat{J}_z \rangle_0$
    denote the expectation values obtained when the protocol is
    initialized in the spin-motion squeezed state and in the ground
    state, respectively.}
    \label{fig:Jzexp}
\end{figure}

The expectation value of $\hat{J}_z$ is given by
\begin{align}
    \langle \hat{J}_z \rangle &\simeq \varphi \sin \vartheta \left[
    \sum_{n=0}^{J} \vert c_{2n} \vert^2 ( - J + 2n ) + J \sum_{n=1}^\infty
    \vert c_{2n+2J} \vert^2 \right] \\
    &= -\varphi \sin \vartheta \left[ J - \frac{z}{1-z} +
    \frac{2}{\sqrt{\pi}} \frac{z^{J+1}}{1-z} \frac{\Gamma
    (J+3/2)}{\Gamma (J+2)} {}_2 F_1 (1/2, J, J+2; z) \right]
    \label{eq:Jzexp}
\end{align}
up to the linear order of $\varphi$.
We compare this result with the case in which the ground state
$\Ket{-J, 0}$ is used in place of $\Ket{{\rm init}}$, for which $\langle
\hat{J}_z \rangle_0 \simeq \varphi J$~\cite{Ito:2025mgm}.
The absolute value of the ratio $\langle \hat{J}_z \rangle / \langle
\hat{J}_z \rangle_0$ is plotted in Fig.~\ref{fig:Jzexp} for $r = 1.0$
(blue), $r = 1.5$ (green), and $r = 2.0$ (red), where, for the purpose
of illustration, the angle $\vartheta$ is fixed to $\pi / 2$.
The scaling of the expectation value with respect to $N = 2 J$ is
improved for sufficiently large $N$, and the ratio approaches unity as
$N \to \infty$.
Although the values are smaller than unity, this $\mathcal{O}(1)$
suppression is compensated in the sensitivity by the variance of
$\hat{J}_z$.

\begin{figure}[t!]
    \centering
    \includegraphics[width=0.6\linewidth]{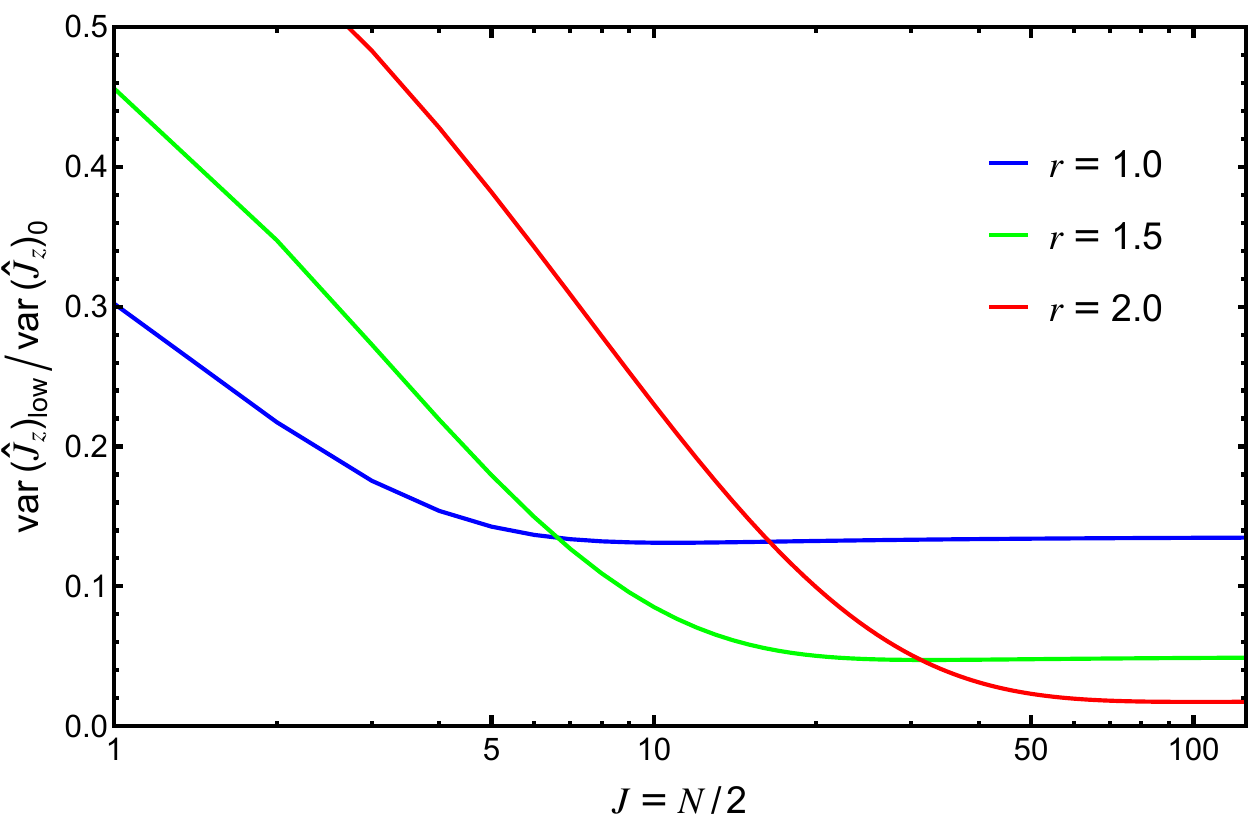}
    \caption{Ratios ${\rm var} (\hat{J}_z)_{\rm low} / {\rm var}
    (\hat{J}_z)_0$ as functions of $N$ for $r = 1.0$ (blue), $r = 1.5$
    (green), and $r = 2.0$ (red), with $\vartheta = \pi / 2$.}
    \label{fig:Jzvar}
\end{figure}

The single-measurement sensitivity $\Delta \varphi$ is defined by
\begin{equation}
    \frac{\vert \varphi \vert}{\Delta \varphi} = \frac{\vert
    \langle \hat{J}_z \rangle \vert}{\sqrt{{\rm var} (\hat{J}_z)}} \simeq
    \frac{\vert \langle \hat{J}_z \rangle \vert}{\sqrt{\langle \hat{J}_z^2
    \rangle}} ,
    \label{eq:sensitivity}
\end{equation}
where the expectation value of $\hat{J}_z^2$ is given by
\begin{align}
    \langle \hat{J}_z^2 \rangle &\simeq \frac{1}{2} \sum_{n=0}^J
    \vert c_{2n} \vert^2 \left( J (J+1) - (-J + 2n)^2 \right) +
    \frac{J}{2} \sum_{n=1}^\infty \vert c_{2n+2J} \vert^2 \notag \\
    &\qquad + \frac{1}{2} \, {\rm Re} \, \left[ e^{-2i\vartheta}
    \sum_{n=0}^J c_{2n+2}^* c_{2n} \sqrt{(2J-2n)(2J-2n-1)(2n+1)(2n+2)}
    \right] ,
\end{align}
neglecting higher-order contributions in $\varphi$.
To gain qualitative insight into the $N$-dependence of ${\rm var}
(\hat{J}_z)$, we consider a lower bound of the variance,
\begin{align}
    {\rm var} (\hat{J}_z)_{\rm low} &= \frac{1}{2} \sum_{n=0}^J
    \vert c_{2n} \vert^2 \left( J (J+1) - (-J + 2n)^2 \right) +
    \frac{J}{2} \sum_{n=1}^\infty \vert c_{2n+2J} \vert^2 \notag \\
    &\qquad - \frac{1}{2} \sum_{n=0}^J \vert c_{2n+2} c_{2n} \vert
    \sqrt{(2J-2n)(2J-2n-1)(2n+1)(2n+2)} ,
\end{align}
by comparison with the variance in the ground-state case,
${\rm var} (\hat{J}_z)_0 = J / 2$.
The ratio ${\rm var} (\hat{J}_z)_{\rm low} / {\rm var} (\hat{J}_z)_0$ is
plotted in Fig.~\ref{fig:Jzvar} for $r = 1.0$ (blue), $r = 1.5$ (green),
and $r = 2.0$ (red).
Although this estimate underestimates the exact values, it captures the
qualitative behavior observed in the numerical results presented in
Sec.~\ref{sec:sensitivity}.
The adiabatically prepared spin-motion squeezed state reduces the
relevant spin variance through squeezing.
As $N$ increases, the protocol becomes advantageous in terms
of the variance of $\hat{J}_z$, which determines the statistical
uncertainty of the measurement, while the ratio exhibits saturation as
$N \to \infty$.

\section{Physics targets}

In this section, we review the target signals considered in this work,
namely wave-like dark matter and high-frequency gravitational waves.
We focus on representative candidates for wave-like dark matter, including
axion-like particles and dark photons, which can induce weak electric
fields that resonantly excite the center-of-mass mode of the ion crystal.
We also discuss high-frequency gravitational waves and their coupling to
the crystal through both indirect and direct mechanisms.
The former arises from graviton-photon conversion in the presence of a
background magnetic field and can likewise excite the center-of-mass mode.
In contrast, the direct coupling allows for the excitation of the
parity-odd modes of the ion crystal.
This feature is unique to gravitational waves and enables their
discrimination from wave-like dark matter signals.


\subsection{Dark matter}

Dark matter with a mass $m_{\rm DM} = \mathcal{O} ({\rm neV})$ behaves as
a classical wave, since the occupation number within a volume of order
the de Broglie wavelength is much larger than unity.
Furthermore, the dark matter distribution is effectively uniform over
this length scale, which is typically much larger than the size of an ion
trap system.
Consequently, the spatial variation of the classical field can be
neglected.
For the case of wave-like dark matter, the dark matter field $\Phi$ can
be expressed as
\begin{equation}
    \Phi ({\bm x}, t) = \Phi_0 \cos (m_{\rm DM} t - \phi) .
\end{equation}
The phase $\phi$ of the field is randomly distributed but remains
approximately constant over the coherence time $T_{\rm DM} = 2 \pi /
m_{\rm DM} v_{\rm DM}^2 \simeq 0.4~{\rm s} \times (10~{\rm neV} /
m_{\rm DM})$, where $v_{\rm DM} \sim
10^{-3}$~\cite{mcmillan2010uncertainty,bovy2009galactic} is the relative
speed of the dark matter.
Since dark matter is non-relativistic, $v_{\rm DM} \ll 1$, the
amplitude $\Phi_0$ is related to the local dark matter density as
$\rho_{\rm DM} = m_{\rm DM}^2 \Phi_0^2 / 2 \sim
0.45$~GeV~cm$^{-3}$~\cite{deSalas:2020hbh,soding2025local}.
We focus on two benchmark models, the axion-like particle and the dark
photon, which can induce a classical electric field that resonantly
excites the center-of-mass mode ($k = 1$) of the ion crystal.

The axion-like particle $\Phi = a$ is a pseudo-scalar field which couples
with photons as
\begin{equation}
    \mathcal{L} = - \frac{1}{4} g_{a \gamma} a F^{\mu\nu}
    \tilde{F}_{\mu\nu} ,
\end{equation}
where $F^{\mu\nu}$ is the electromagnetic field tensor, and
$\tilde{F}^{\mu\nu}$ is its dual.
Under a static magnetic field $B_z$ of radius $R$, the axion field is
converted into an electric field
\begin{equation}
    E_z = \epsilon_a \sqrt{2 \rho_{\rm DM}} \sin (m_{\rm DM} t - \phi)
\end{equation}
along the magnetic field, where
\begin{equation}
    \epsilon_a = \frac{B_z}{2} \frac{g_{a\gamma}}{m_{\rm DM}} (m_{\rm DM}
    R)^2 \left[ \left( \log \frac{m_{\rm DM} R}{2} + \gamma - \frac{1}{2}
    \right)^2 + \left( \frac{\pi}{2} \right)^2 \right]^{1/2}
    \label{epa}
\end{equation}
is the conversion efficiency for $m_{\rm DM} R \ll 1$ with $\gamma =
0.5772 \dots$ is the Euler--Mascheroni constant~\cite{Ouellet:2018nfr}.
When the frequency $\omega_z$ matches the dark matter mass $m_{\rm DM}$,
this induced electric field excites the center-of-mass mode of the ion
crystal via the interaction Hamiltonian~\eqref{eq:int} with
\begin{equation}
    \alpha = -i e^{-i\phi} \frac{e \epsilon_a}{2}
    \sqrt{\frac{N \rho_{\rm DM}}{m_{\rm DM} \omega_z}}
    \label{eq:alpha}
\end{equation}
under the rotating wave approximation.

The dark photon is a massive vector boson associated with
an extra ${\rm U}(1)$ gauge group besides the Standard Model.
It has very weak interactions with the Standard Model particles via the
kinetic mixing with the photon,
\begin{equation}
    \mathcal{L} = \frac{1}{2} \epsilon F^{\mu\nu} F'_{\mu\nu} ,
\end{equation}
where $F'^{\mu\nu}$ is the field strength tensor of the dark photon field.
The electric field component of the dark photon dark matter is given by
${\bm E} = {\bm E}_0 \sin (m_{\rm DM} t - \phi)$, with the amplitude
related to the local dark matter density, $\rho_{\rm DM} = {\bm E}_0^2 / 2$.
We focus on the $z$-component
\begin{equation}
    E_z = \epsilon_{\rm DP} \sqrt{2 \rho_{\rm DM}}
    \sin (m_{\rm DM} t - \phi)
\end{equation}
with $\epsilon_{\rm DP} \simeq \epsilon \cos \theta (m_{\rm DM}
R)^2$~\cite{Chaudhuri:2014dla}.
The polarization angle $\theta$ is assumed to be randomly distributed.
The interaction of the dark photon dark matter is described by
Eqs.~\eqref{eq:int} and \eqref{eq:alpha} with replacement of $\epsilon_a$
with $\epsilon_{\rm DP}$.


\subsection{Gravitational waves}

To correctly describe the effect of gravitational waves on the ion
crystal, it is essential to adopt an appropriate coordinate system.
The most convenient choice is the proper detector frame,
whose origin follows the geodesic of the center
of mass of the ion crystal~\cite{Manasse:1963zz,Ni:1978zz}.
In this frame, the origin is chosen to coincide with the freely falling
center of the crystal, such that the coordinate values faithfully
represent the positions of individual ions.
Within this formulation, both the center-of-mass mode ($k = 1$) and the
parity-odd modes ($k = 2$, $3$) can be resonantly excited by gravitational
waves.

We consider gravitational waves propagating on a Minkowski background,
with the metric expressed as $g_{\mu\nu} = \eta_{\mu\nu} + h_{\mu\nu}$,
where $\eta_{\mu\nu}$ is the Minkowski metric and $h_{\mu\nu}$ denotes the
transverse-traceless perturbation.
We model the gravitational wave as a plane wave,
\begin{equation}
    h_{ij} ({\bm x}, t) = h_+ e_{ij}^+ \cos (\omega t - {\bm k} \cdot
    {\bm x} + \phi_+) + h_\times e_{ij}^\times \cos (\omega t - {\bm k}
    \cdot {\bm x} + \phi_\times) ,
\end{equation}
where $\omega = 2 \pi f$ is the angular frequency, ${\bm k}$ is the
wavevector, and $h_+$, $h_\times$ together with $\phi_+$, $\phi_\times$
denote the amplitudes and phases of the two polarization modes,
respectively.
The corresponding polarization tensors are given by
\begin{equation}
    e_{ij}^+ = \frac{1}{\sqrt{2}}
    \begin{pmatrix}
        \cos^2 \theta & 0 & - \sin \theta \cos \theta \\
        0 & -1 & 0 \\
        - \sin \theta \cos \theta & 0 & \sin^2 \theta
    \end{pmatrix} , \quad e_{ij}^\times = \frac{1}{\sqrt{2}}
    \begin{pmatrix}
        0 & \cos \theta & 0 \\ \cos \theta & 0 & -\sin \theta \\
        0 & -\sin \theta & 0
    \end{pmatrix} ,
\end{equation}
where the $+$ polarization corresponds to a deformation along the
$y$-direction, and $\theta$ denotes the propagation angle in the
$xz$-plane.
Hereafter, we assume an unpolarized gravitational wave, such that
$h_+ = h_\times \equiv h_0$ and $\phi_+ = \phi_\times \equiv \phi$,
and randomness of the polarization $\theta$ and the phase $\phi$
for simplicity.

Gravitational waves are known to convert into electromagnetic waves in the
presence of a background magnetic field.
The center-of-mass mode can be excited indirectly via the generated
electromagnetic field.
This graviton-photon conversion can be described by an effective
interaction of the form
\begin{equation}
    \mathcal{L} = j^\mu_{\rm eff} A_{\mu} , \quad j^\mu_{\rm eff} =
    \partial_\nu \left( h^\mu_{\,\lambda} \bar{F}^{\lambda\nu} -
    h^\nu_{\,\lambda} \bar{F}^{\lambda\mu} \right) - \frac{1}{2}
    \left( \partial_\nu h^\lambda_{\,\lambda} \right) \bar{F}^{\nu\mu} ,
\end{equation}
where $\bar{F}^{\mu\nu} = \partial^\mu \bar{A}^\nu - \partial^\nu
\bar{A}^\mu$ denotes the field strength tensor associated with the
background electromagnetic field $\bar{A}^\mu$~\cite{Ratzinger:2024spd}.
The metric $h_{\mu\nu}$ in the effective current is evaluated in
Fermi normal coordinates~\cite{Ratzinger:2024spd}.
Note that the induced electric field is independent of the details of
the ion trap system apart from the boundary effects.
We assume that the electrodes and other metallic shielding are placed
outside the magnetic field, which allows us to neglect the boundary
effects.
Under the long-wavelength approximation $\omega R \ll 1$, this interaction
induces an electric field along the axial direction of the form
\begin{equation}
    E_z \simeq \frac{h_0}{\sqrt{2}} B_z (\omega R)^2 f(L/2R) \sin^2
    \theta \cos (\omega t + \phi) ,
\end{equation}
where $B_z$ is a cylindrically symmetric background magnetic field of
radius $R$ and height $L$, and $f(x) = \log (x + \sqrt{1 + x^2})$ is
determined by the magnetic field geometry~\cite{Takai:2025cyy}.
The electric-field contribution is negligible compared with the
magnetic-field contribution, since the trapping electric field is
$\sim 1~{\rm V}/{\rm cm} \sim 10^{-4}~{\rm eV}^2$, whereas the
magnetic field is $\sim 1~{\rm T} \sim 10^2~{\rm eV}^2$.

In the non-relativistic limit, the interaction between the gravitational
field and an ion is described by $\frac{1}{2} T^{\mu\nu} \delta g_{\mu\nu}
\simeq \frac{1}{2} T^{00} \delta g_{00}$, where $T^{\mu\nu}$ is the
energy-momentum tensor of the ion, and $\delta g_{\mu\nu}$ is the metric
perturbation.
In the proper detector frame, this leads to the interaction Hamiltonian
$H = \frac{1}{2} m_{\rm ion} R_{0k0l} x^k x^l$, where
$R_{\mu\nu\rho\sigma}$ is the Riemann tensor evaluated at the origin $x^i
= 0$~\cite{Ito:2020wxi}.
Here, the transverse-traceless gauge is employed to calculate
$R_{0k0l}$, because the Riemann tensor is gauge invariant at linear order
in the metric perturbation.
We obtain the effective interaction Hamiltonian describing the coupling
between a plane gravitational wave and the ion crystal as
Eq.~\eqref{eq:int} with
\begin{equation}
    \alpha \simeq - \frac{\sqrt{N m_{\rm ion}} \omega_2^{3/2} R_{\rm cry}}{8
    \zeta_{11} \sqrt{\zeta_{11}^2 - 1}} h_0 \sin \theta (\cos \theta - i)
    e^{i \phi} ,
\end{equation}
where $\zeta_{11} = 1.841 \dots$ is the first zero of the first Bessel
function and $R_{\rm cry}$ denotes the radius of the
ion crystal, for the parity-odd mode $\hat{a} = (\hat{a}_2 + i \hat{a}_3)
/ \sqrt{2}$~\cite{Ito:2025mgm}.
Henceforth, we consider monochromatic gravitational waves
for simplicity.


\section{Sensitivity}
\label{sec:sensitivity}

In this section, we present results obtained from numerical calculations
of the time evolution governed by the Tavis--Cummings Hamiltonian.
For the adiabatic transition employed in the following simulations, the
parameters are chosen as $\zeta (t) = \zeta_0 \sin (\gamma t/2)$,
$\lambda(t) = \lambda_0 \cos^2 (\gamma t/2)$, $\zeta_0 / 2\pi = 2.5$~kHz,
$\lambda_0 / 2\pi = 1.5$~kHz, $- t_i = t_f = \pi / \gamma$, and $\gamma
/ 2\pi = 2.75$~kHz.
The Hilbert space is truncated by imposing $\vert c_{2n+2J} \vert =
10^{-3}$.
The duration of the protocol is $T = 0.04$~s $< T_{\rm DM} (m_{\rm DM}
= 10$~neV), the optical dipole force coupling is $g / 2\pi = 3.9$~kHz,
and the duration of a single optical dipole force operation is $\tau =
0.3$~ms. 
In this setup, the rotation angle satisfies $\vartheta \simeq \pi/2$,
such that $\sin \vartheta \simeq 1$.

\begin{figure}[t!]
    \begin{minipage}[t]{0.49\linewidth}
        \includegraphics[width=\linewidth]{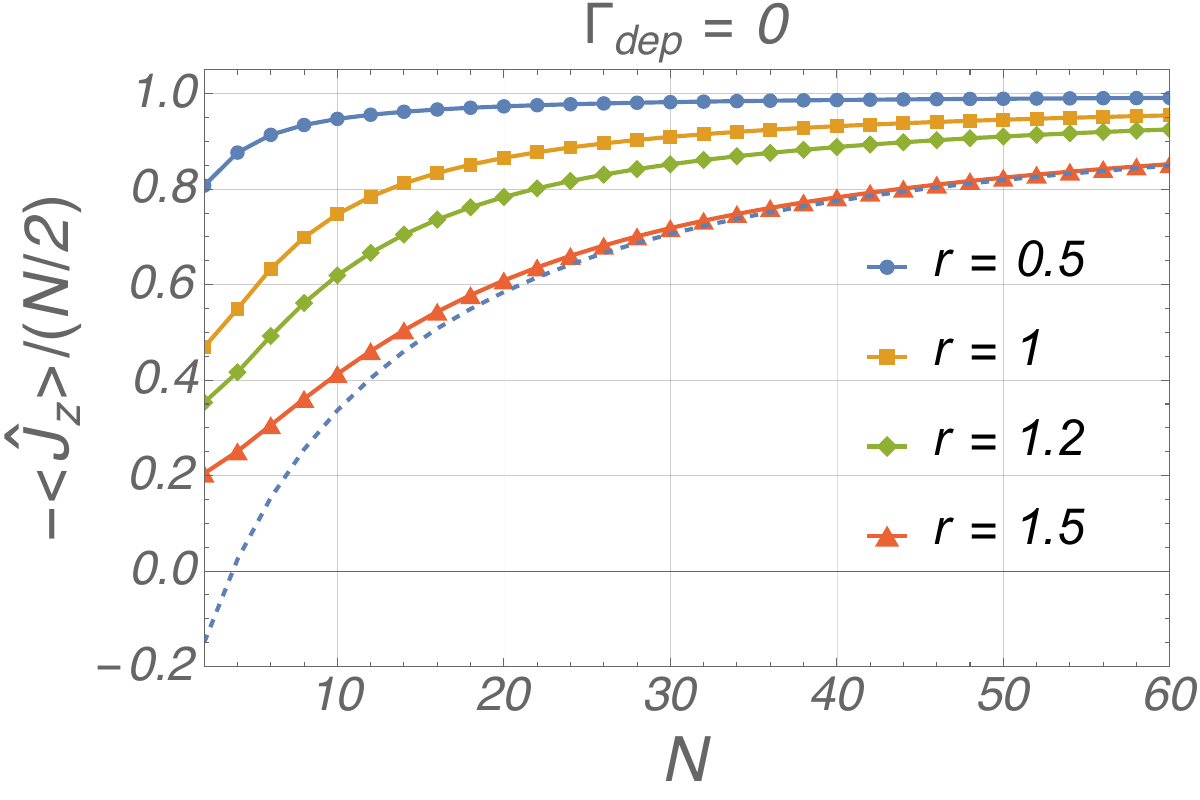}
    \end{minipage}
    \hspace{0.01\linewidth}
    \begin{minipage}[t]{0.49\linewidth}
        \includegraphics[width=\linewidth]{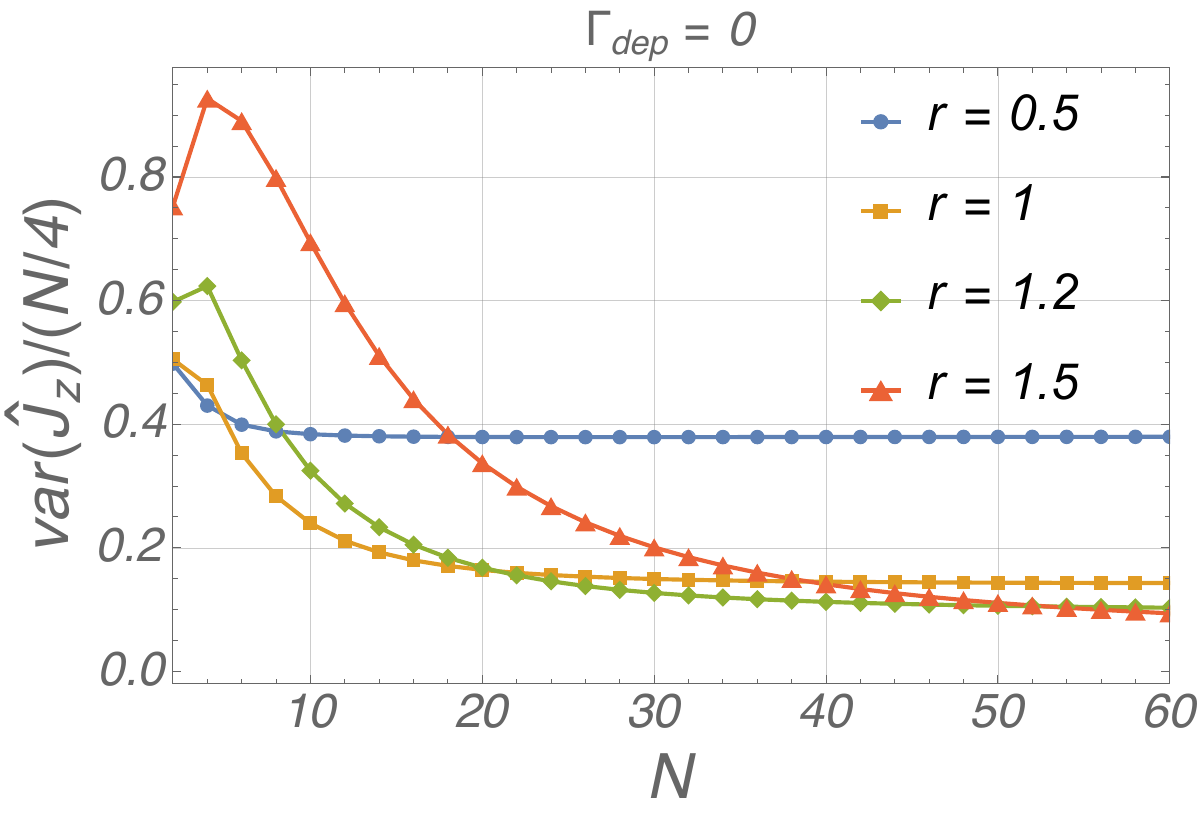}
    \end{minipage}
    \caption{Numerical results of the expectation value (left)
    and the variance (right) of $\hat{J}_z$ as a function of $N$ for
    $r = 0.5$ (blue), $r = 1.0$ (orange), $r = 1.2$ (green),
    and $r = 1.5$ (red).
    The results are compared with those obtained when the protocol is
    initialized in the ground state instead of the spin-motion squeezed
    state.
    The dashed curve in the left panel shows the result obtained from
    Eq.~\eqref{eq:Jzexp} for $r = 1.5$.}
    \label{fig:S_z}
\end{figure}

We first examine the expectation value and variance of $\hat{J}_z$.
Fig.~\ref{fig:S_z} shows these quantities in comparison with the case
where the protocol is initialized in the ground state, for which
$\langle \hat{J}_z \rangle_0 = N / 2$ and ${\rm var} (\hat{J}_z)_0 =
N / 4$.  We plot $-\langle\hat{J}_z\rangle$ because the response is negative.
In the left panel, the analytical result based on Eq.~\eqref{eq:Jzexp}
is shown by the dashed curve.
The deviation between analytical and numerical results at small $N$
arises from incomplete adiabaticity due to insufficient level crossings,
which breaks the conservation of $\hat{\mathcal{N}}$ and is not captured
by Eq.~\eqref{eq:init}.
This discrepancy becomes negligible when the transition time is
increased by an order of magnitude.
Both panels reproduce the qualitative behavior discussed in
Sec.~\ref{sec:protocol} for sufficiently large $N$.

\begin{figure}[t!]
    \begin{minipage}[t]{0.49\linewidth}
        \includegraphics[width=\linewidth]{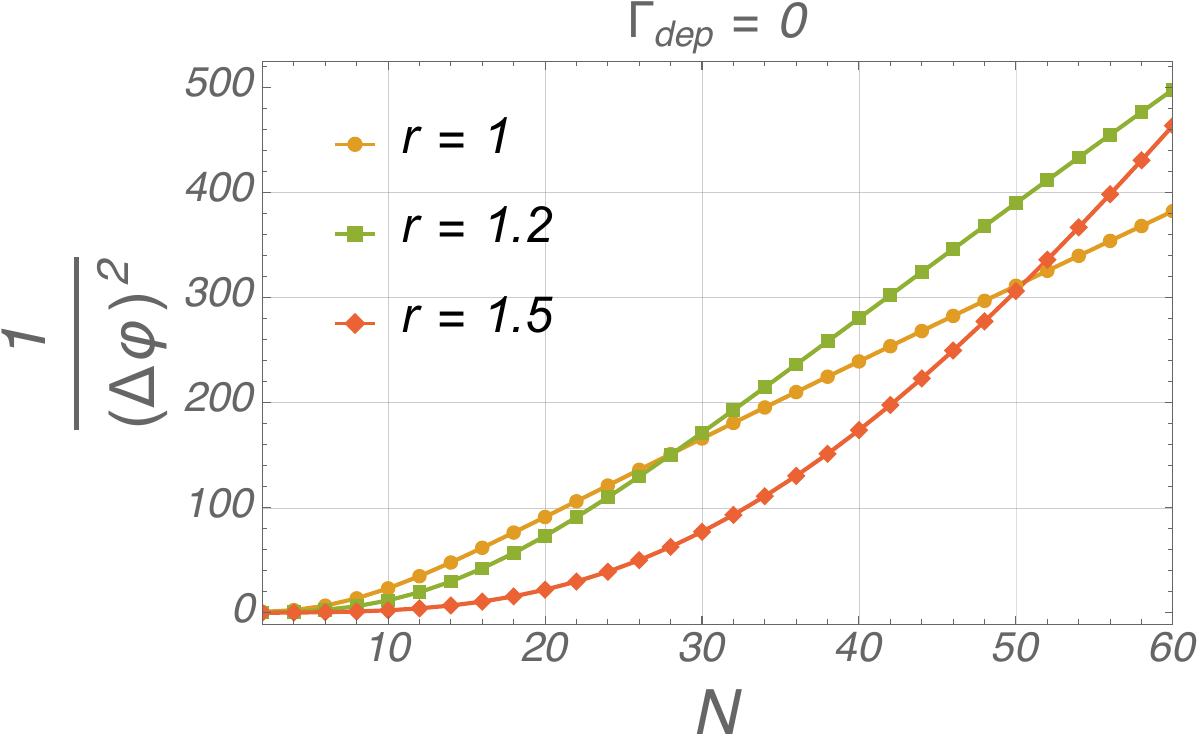}
    \end{minipage}
    \hspace{0.01\linewidth}
    \begin{minipage}[t]{0.49\linewidth}
        \includegraphics[width=\linewidth]{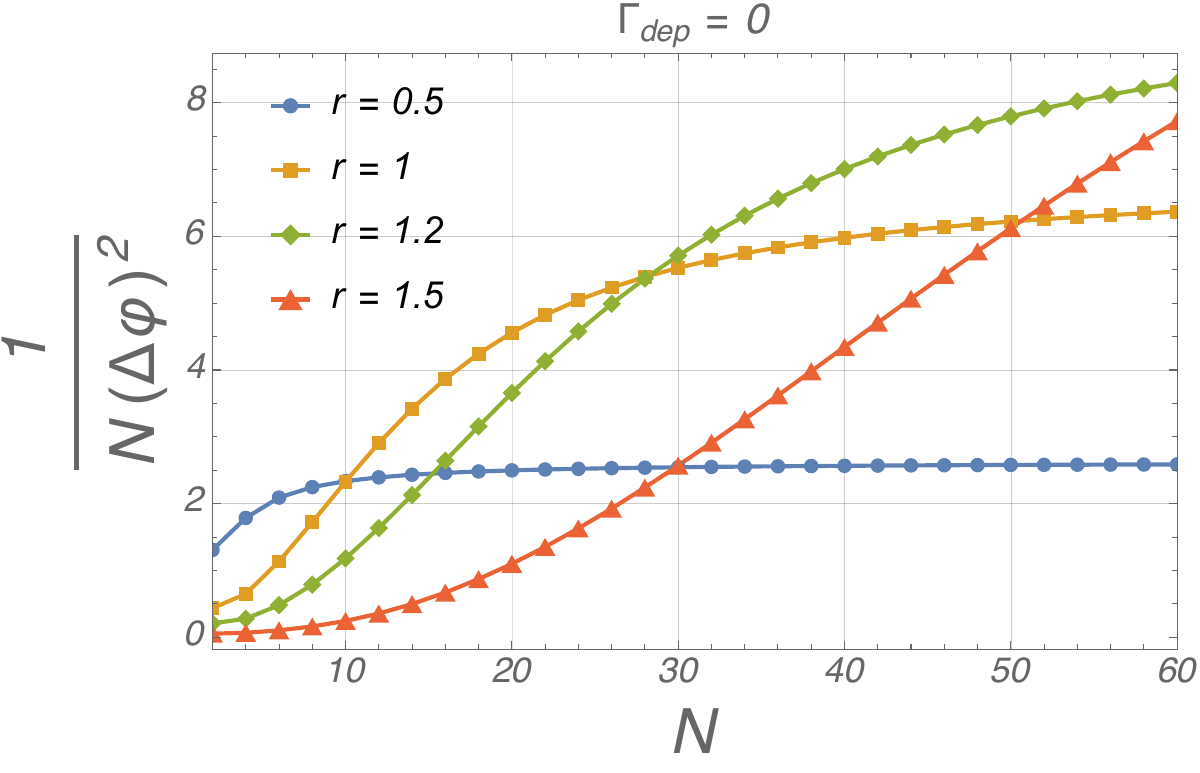}
    \end{minipage}
    \caption{Sensitivities $(\Delta\varphi)^{-2}$ as a function of $N$
    for $r = 0.5$ (blue), $r = 1.0$ (orange), $r = 1.2$ (green), and
    $r = 1.5$ (red).
    The right panel shows the sensitivity relative to that obtained with
    a non-squeezed initial state;
    values greater than unity indicate an enhancement due to the
    spin-motion squeezed state.}
    \label{fig:sensitivity}
\end{figure}

Next, the sensitivity $\Delta \varphi$, defined in
Eq.~\eqref{eq:sensitivity}, is shown in Fig.~\ref{fig:sensitivity}
for several values of $r$.
The right panel compares the sensitivity with that obtained when the
protocol is initialized in the ground state.
For sufficiently large $N$, the inverse sensitivity normalized by the
ground-state result exceeds unity, indicating an improvement over the
non-squeezed case.

\begin{figure}[t!]
    \centering
    \includegraphics[width=0.6\linewidth]{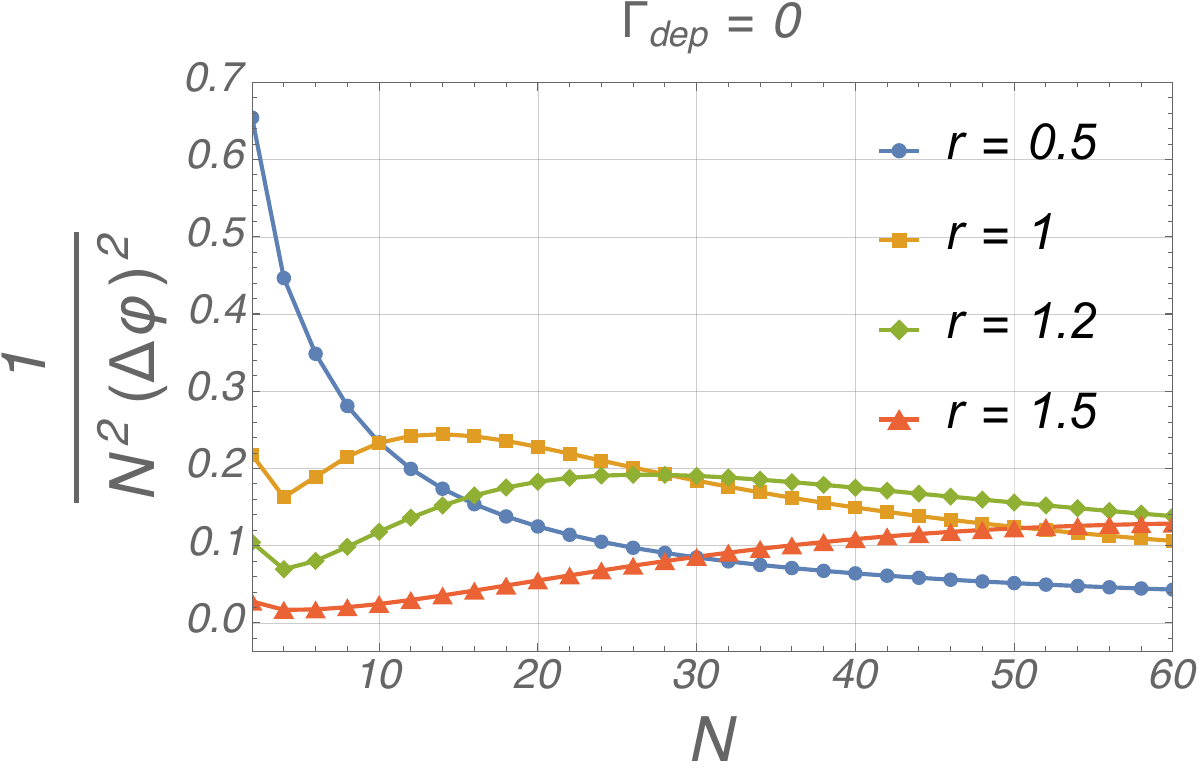}
    \caption{Sensitivities $(N \Delta\varphi)^{-2}$ as a function of $N$
    for $r = 0.5$ (blue), $r = 1.0$ (orange), $r = 1.2$ (green), and
    $r = 1.5$ (red).
    On the plateau, the sensitivity follows the Heisenberg scaling
    $(\Delta\varphi)^2 \propto N^{-2}$, whereas in the regime where the
    curves increase, it exhibits super-Heisenberg scaling $\propto
    N^{-2k}$ with $k > 1$.}
    \label{fig:xi_SH}
\end{figure}

Fig.~\ref{fig:xi_SH} shows the sensitivity in comparison with the
Heisenberg scaling $(\Delta\varphi)^2 \propto N^{-2}$.
A positive slope in this figure indicates scaling faster than $N^2$,
corresponding to super-Heisenberg behavior.
We observe such scaling over a range of $N$ that depends on the choice
of $r$, and larger values of $r$ extend this regime to higher $N$.

\begin{figure}[t!]
    \centering
    \includegraphics[width=0.6\linewidth]{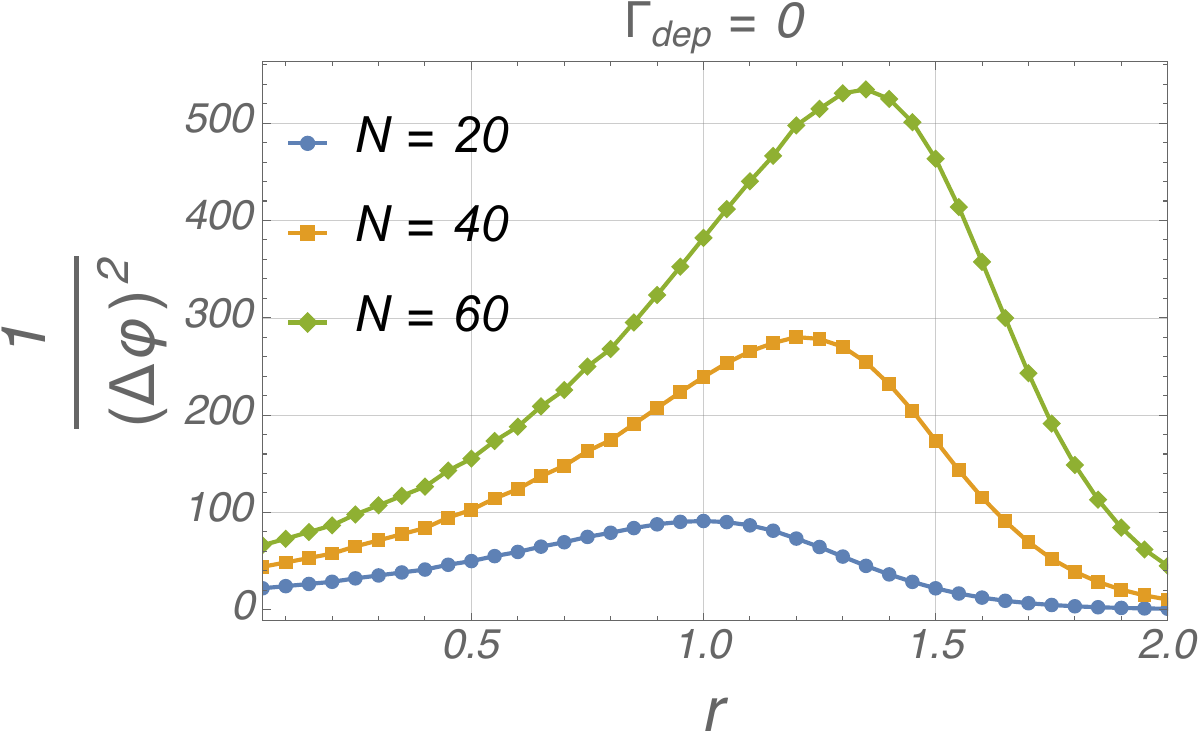}
    \caption{Sensitivities $(\Delta\varphi)^{-2}$ as a function of the
    squeezing parameter $r$ for $N = 20$ (blue), $N = 40$ (orange), and
    $N = 60$ (green).}
    \label{fig:Dphi_r}
\end{figure}

Fig.~\ref{fig:Dphi_r} shows the dependence of the sensitivity
$(\Delta\varphi)^{-2}$ on the squeezing parameter $r$.
A peak structure is observed, whose location depends on $N$, indicating
that the optimal value of $r$ should be tuned for a given system size.

\begin{figure}[t]
    \begin{minipage}[t]{0.49\linewidth}
        \includegraphics[width=\linewidth]{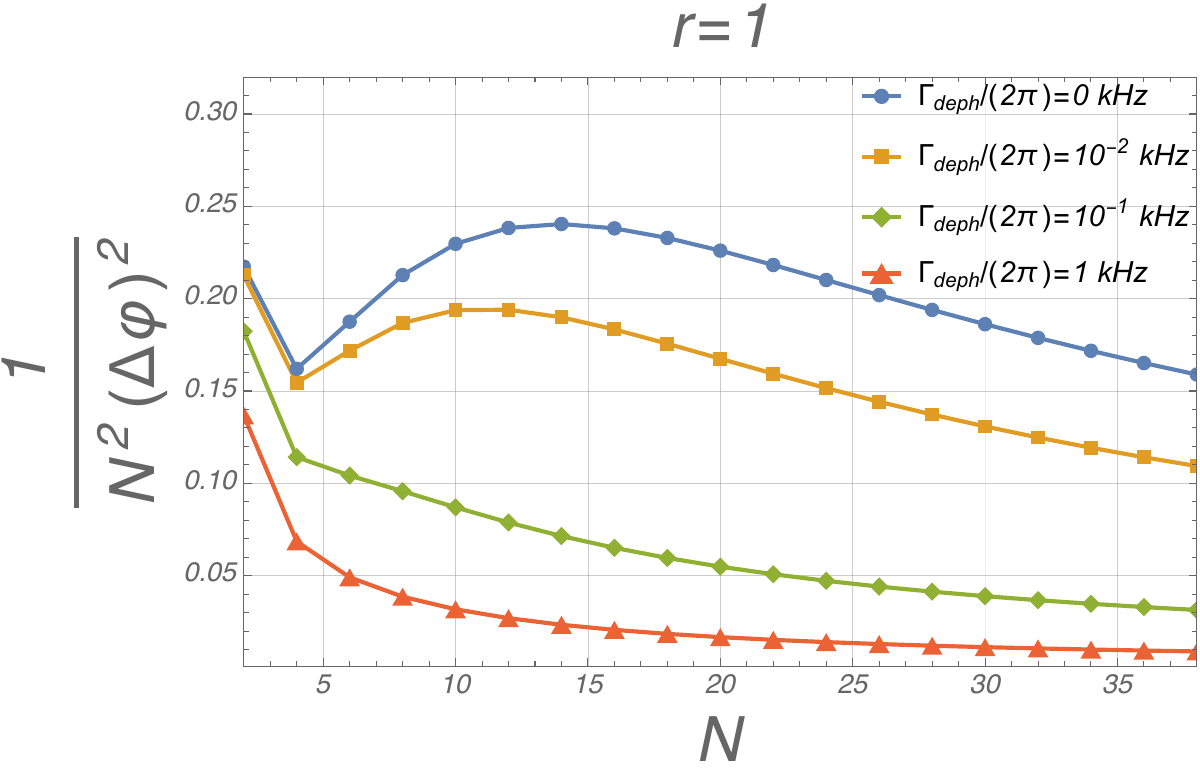}
    \end{minipage}
    \hspace{0.01\linewidth}
    \begin{minipage}[t]{0.49\linewidth}
        \includegraphics[width=\linewidth]{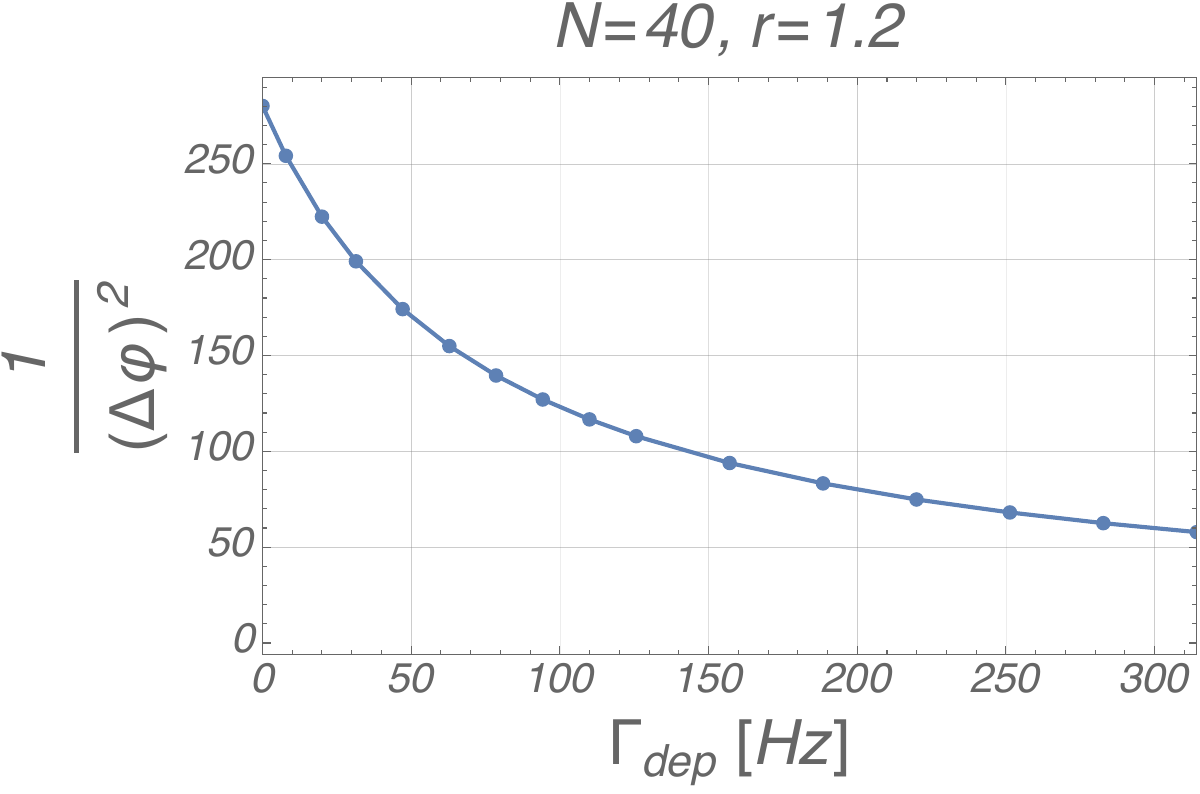}
    \end{minipage}
    \caption{Sensitivities $(N \Delta\varphi)^{-2}$ as functions of $N$
    (left) for $\Gamma_{\rm dep} / 2\pi = 0$ (blue), 0.01~kHz (orange),
    0.1~kHz (green), 1~kHz (red) with $r = 1$, and as a function
    of $\Gamma_{\rm dep}$ (right) for $N = 40$ and $r = 1.2$.}
    \label{fig:deph} 
\end{figure}

In a realistic setup, spin dephasing during the adiabatic transition
must be taken into account.
This effect is incorporated by solving the master equation for the
density matrix
\begin{equation}
    \frac{\partial \hat{\rho}}{\partial t} = -i \left[ \hat{H}_{\rm TC},
    \hat{\rho} \right] + \Gamma_{\rm dep} \left( \hat{J}_z \hat{\rho}
    \hat{J}_z - \frac{1}{2} \left\{ \hat{J}_z^2, \hat{\rho} \right\}
    \right) .
\end{equation}
The numerical results are shown in Fig.~\ref{fig:deph}.
In the left panel, the sensitivity follows Heisenberg scaling
$(\Delta\varphi)^2 \propto N^{-2}$ on a plateau, and exhibits
super-Heisenberg scaling $\propto N^{-2k}$ with $k > 1$ in the regime
where the curves increase.
We find that the super-Heisenberg region is reduced by dephasing, and the
sensitivity degrades exponentially with increasing $\Gamma_{\rm dep}$.
Decoherence during the optical dipole force operations is also included
via an exponential suppression factor $e^{-2\Gamma_{\rm dec} \tau}$
applied to $(\Delta\varphi)^{-2}$~\cite{Gilmore:2021qqo}.
In the following estimates, we assume $\Gamma_{\rm dep} =
\Gamma_{\rm dec} = 100$~Hz, and neglect laser detuning and thermal
noise for simplicity.
We note that thermal noise slightly affects the effective
super-Heisenberg regime for heating rates of $\mathcal{O}
(1)$~ms$^{-1}$~\cite{Pavlov:2024uxs}.

\begin{figure}[t]
    \begin{minipage}[t]{0.49\linewidth}
        \includegraphics[width=\linewidth]{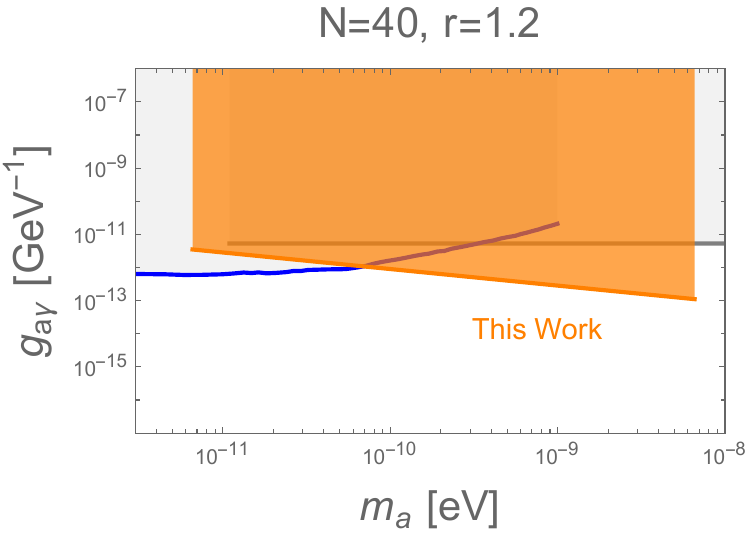}
    \end{minipage}
    \hspace{0.01\linewidth}
    \begin{minipage}[t]{0.49\linewidth}
        \includegraphics[width=\linewidth]{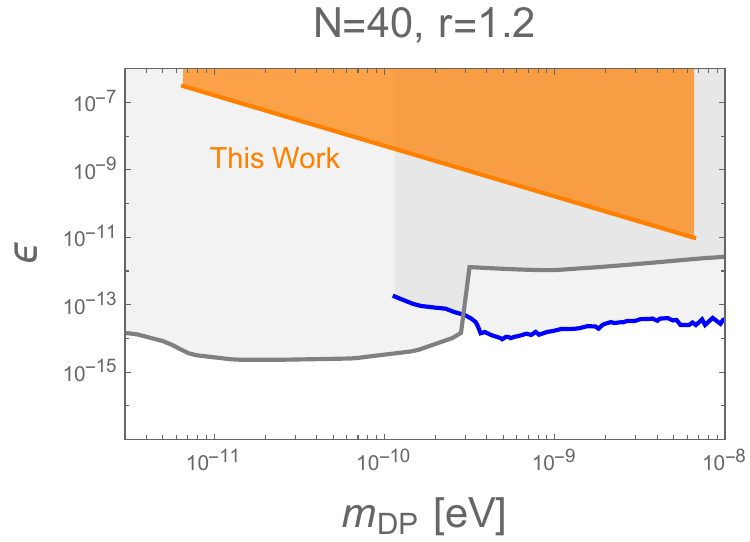}
    \end{minipage}
    \caption{Sensitivity to the axion-photon coupling (left) and the
    kinetic mixing of the dark photon (right) as functions of the dark
    matter mass, $m_{\rm DM} = m_a$ or $m_{\rm DP}$, obtained by
    scanning over the mass range during a one-year measurement.
    The blue and gray region in the left panel are excluded by the
    M82~\cite{Ning:2024eky} and by the magnetic white
    dwarfs~\cite{Dessert:2022yqq}, respectively.
    The blue and gray region in the right panel are constrained by the
    spacecraft, Parker Solar Probe~\cite{An:2024wmc}, and by the heating
    of cosmic gases~\cite{McDermott:2019lch}, respectively.}
    \label{fig:DM}
\end{figure}

Finally, we estimate the sensitivities to dark matter and gravitational
waves.
The sensitivities at 68\% confidence level (corresponding to
Eq.~\eqref{eq:sensitivity})
to the axion-photon coupling $g_{a\gamma}$ and the kinetic mixing
parameter of the dark photon $\epsilon$ are given by
\begin{align}
    g_{a\gamma} \simeq 1.2 \times 10^{-14}~{\rm GeV}^{-1} &\times \left(
    \frac{\Delta \varphi}{0.0081} \right) \left(
    \frac{m_{\rm ion}}{8.3~{\rm GeV}} \right)^{1/2} \left(
    \frac{m_{\rm DM}}{1~{\rm neV}} \right)^{-1/2} \notag \\
    &\qquad \times \left( \frac{B_z}{1~{\rm T}} \right)^{-1} \left(
    \frac{R}{3~{\rm m}} \right)^{-2} \left(
    \frac{T_{\rm tot}}{1~{\rm day}} \right)^{-1/2}
\end{align}
and
\begin{equation}
    \epsilon = 1.2 \times 10^{-11} \times \left( \frac{\Delta
    \varphi}{0.0081} \right) \left( \frac{m_{\rm ion}}{8.3~{\rm GeV}}
    \right)^{1/2} \left( \frac{m_{\rm DM}}{1~{\rm neV}} \right)^{-3/2}
    \left( \frac{R}{3~{\rm m}} \right)^{-2} \left(
    \frac{T_{\rm tot}}{1~{\rm day}} \right)^{-1/2} ,
\end{equation}
respectively, for $N = 40$ and $r = 1.2$\footnote{The sensitivity for
the dark photon improves by $\sim 10^4$ under free boundary conditions.}.
We assume that the measurement is repeated $N_{\rm rep} = T_{\rm tot} /
(T+2\pi/\gamma)$ times.
Fig.~\ref{fig:DM} shows the sensitivities obtained by scanning over the
dark matter mass during a one-year measurement.
The number of bins is taken to be $N_{\rm bin} = Q \log (f_{\rm max} /
f_{\rm min})$ with $Q \sim 10^4$, $f_{\rm max} = 10$~MHz, and
$f_{\rm min} = 10$~kHz~\cite{Gilmore:2021qqo}.
Also shown are existing experimental constraints~\cite{AxionLimits}.
This setup can potentially exclude the currently allowed parameter
regions, and in particular, for axion-like particles, it enables access
to previously unexplored regions even for relatively small $N$ and $r$.
Larger values of $N$ and $r$ further improve the sensitivity.

\begin{figure}[t]
    \begin{minipage}[t]{0.49\linewidth}
        \includegraphics[width=\linewidth]{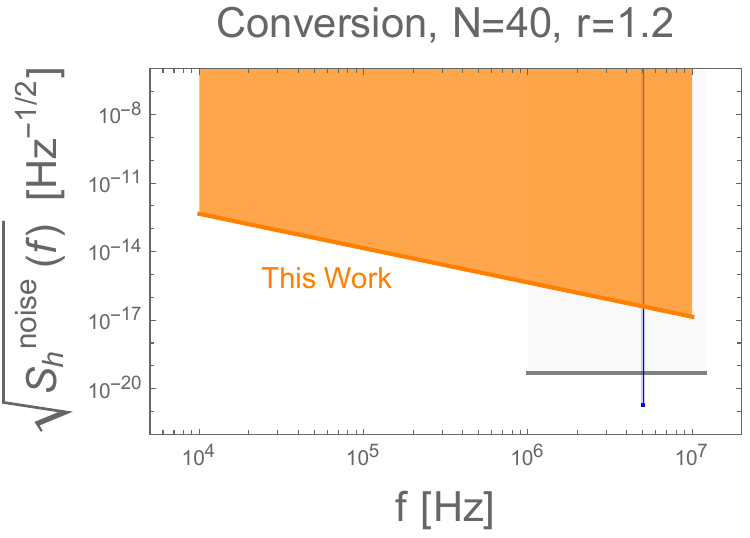}
    \end{minipage}
    \hspace{0.01\linewidth}
    \begin{minipage}[t]{0.49\linewidth}
        \includegraphics[width=\linewidth]{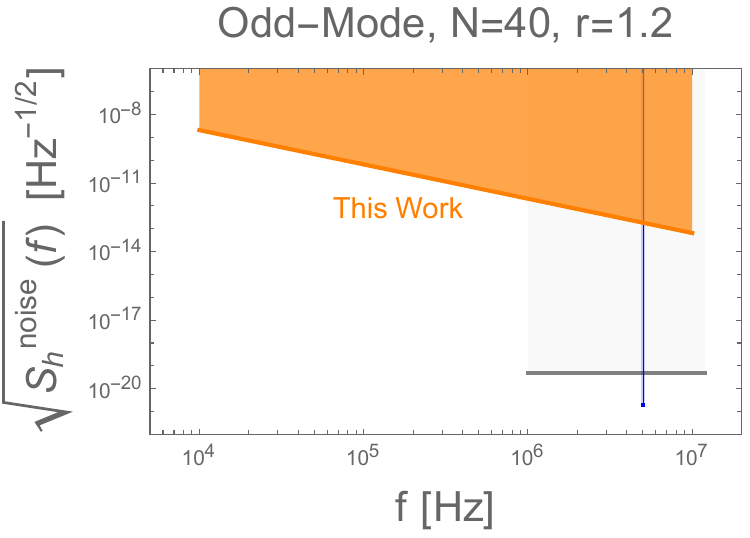}
    \end{minipage}
    \caption{Sensitivities to the noise-equivalent spectral density of
    gravitational waves as functions of frequency, obtained by scanning
    over the frequency range during a one-year observation, via the
    center-of-mass mode (left) and the parity-odd modes (right).
    The gray band and the blue line indicate existing experimental
    constraints from the Fermilab Holometer~\cite{chou2017mhz} and the
    Bulk Acoustic Wave experiments~\cite{Goryachev:2014yra},
    respectively.}
\label{fig:GW}
\end{figure}

The sensitivities at 68\% confidence level to the gravitational wave
strain are given by
\begin{align}
    h_0 (f) \simeq 1.6 \times 10^{-16} &\times \left(
    \frac{\Delta \varphi}{0.0081} \right) \left(
    \frac{m_{\rm ion}}{8.3~{\rm GeV}} \right)^{1/2} \left(
    \frac{f}{1~{\rm MHz}} \right)^{-3/2} \notag \\
    &\qquad \times \left( \frac{R}{3~{\rm m}} \right)^{-2} \left(
    \frac{B_z}{1~{\rm T}}\right)^{-1} \left(
    \frac{T_{\rm tot}}{1~{\rm day}} \right)^{-1/2}
\end{align}
via graviton-photon conversion, where $L=1$ m and 
\begin{equation}
    h_0 (f) \simeq 7.5 \times 10^{-13} \times \left(
    \frac{\Delta \varphi}{0.0081} \right) \left(
    \frac{m_{\rm ion}}{8.3~{\rm GeV}} \right)^{-1/2} \left(
    \frac{f}{1~{\rm MHz}} \right)^{-3/2} \left(
    \frac{R_{\rm cry}}{0.1\, {\rm mm}} \right)^{-1} \left(
    \frac{T_{\rm tot}}{1~{\rm day}} \right)^{-1/2}
\end{equation}
via the parity-odd mode.
Using the same parameters as in the dark matter case, the sensitivity
to the power spectrum density $S_h(f) = T h_0^2 (f)$ at 68\% confidence
level, i.e., the noise-equivalent spectral
density~\cite{Aggarwal:2025noe}, is shown in Fig.~\ref{fig:GW}.
The sensitivities are obtained by scanning over frequency during a
one-year observation with the same binning as in the dark matter
analysis.
The ion crystal enables searches in frequency ranges that have not been
explored by existing experiments, and larger values of $N$ and $r$ can
achieve sensitivities comparable to or exceeding current limits.

\section{Conclusion}

We investigate a quantum-enhanced scheme for the detection of wave-like
dark matter and high-frequency gravitational waves using two-dimensional
ion crystals.
We propose a detection protocol initialized in a spin-motion squeezed
state, adiabatically prepared via motional squeezing and the
Tavis--Cummings Hamiltonian, in which weak external forces are converted
into measurable collective-spin signals.
We show that this protocol improves the signal-to-noise ratio and
enables super-Heisenberg scaling with respect to the number of ions.
We also estimate the corresponding sensitivities to wave-like dark
matter candidates, such as the axion-like particle and the dark photon,
as well as to high-frequency gravitational waves.

At the same time, the enhancement is sensitive to experimental
imperfections.
In particular, we find that dephasing reduces the parameter region in
which super-Heisenberg scaling appears, while decoherence during the
optical dipole force operations degrades the overall sensitivity.
Additional effects neglected in the present analysis, such as finite
temperature, frequency detuning errors, and imperfect adiabaticity,
should be incorporated in a more complete assessment of experimental
performance.
Furthermore, our analysis focuses on the regime $N < 100$ and $r < 2$
due to computational limitations, whereas larger values are advantageous
for probing weaker interactions.
Extending this framework to other ion-trap architectures is an
interesting direction for future work.


\section*{Acknowledgment}

This work was supported by JSPS KAKENHI Grant Numbers
JP23KJ2173 (WN) and JP24KJ1157 (RT).


\bibliography{bibcollection}
\bibliographystyle{modifiedJHEP}


\end{document}